# Acceleration of cerebral blood flow and arterial transit time maps estimation from multiple post-labeling delay arterial spin-labeled MRI via deep learning

Yiran Li[1], Ze Wang[1]

[1] Department of Diagnostic Radiology and Nuclear Medicine, University of Maryland School of Medicine, Baltimore, MD 21201, USA

Correspondence:

Ze Wang, Ph.D.

Department of Diagnostic Radiology and Nuclear Medicine,

University of Maryland School of Medicine

Baltimore, MD 21201, USA

Email: Ze.Wang@som.umaryland.edu

**Abstract.**

**Purpose:** Arterial spin labeling (ASL) perfusion imaging indicates direct and absolute measurement of cerebral blood flow (CBF). Arterial transit time (ATT) is a related physiological parameter reflecting the duration for the labeled spins to reach the brain region of interest. Multiple post-labeling delay (PLDs) can provide robust measures of both CBF and ATT, allowing for optimization of regional CBF modeling based on ATT. The prolonged acquisition time can potentially reduce the quality and accuracy of the CBF and ATT estimation. We proposed a novel network to significantly reduce the number of PLDs with higher signal-to-noise ratio (SNR).

**Method**: CBF and ATT estimations were performed for one PLD and two PLDs separately. Each model was trained independently to learn the nonlinear transformation from perfusion weighted image (PWI) to CBF and ATT images.

**Results**: Both one-PLD and two-PLD models outperformed the conventional method visually on CBF and two-PLD model showed more accurate structure on ATT estimation. The proposed method significantly reduces the number of PLDs from 6 to 2 on ATT and even to single PLD on CBF without sacrificing the SNR.

**Conclusion**: It is feasible to generate CBF and ATT maps with reduced PLDs using deep learning with high quality.

# 1    Introduction

Cerebral blood flow (CBF) is a physiological measure fundamental to brain neurovascular condition and brain function. Arterial spin labeling (ASL) perfusion MRI remains the only non-invasive and non-radioactive technique for measuring cerebral blood flow (CBF) across the whole brain [1]. ASL MRI relies on the magnetically inverted inflowing arterial blood as the endogenous tracer. The labeled arterial blood will be transit to the imaging place and exchange with tissue water and subsequently change the tissue signal. This signal change is in proportion to cerebral perfusion (CBF) and is encoded in the corresponding image acquired in the imaging place. By convention, this image is often called the label (L) image. To remove the background signal, a paired image acquisition is performed using the same spin labeling and imaging sequence after modulating the spin labeling pulses to not invert the arterial spins. This companion image is often named the control (C) image. Perfusion signal is subsequently extracted from the difference between C and L images, which can be converted into the quantitative CBF in a unit of ml/100g/min using a kinetic model [2]. Because arterial spins are often labeled in a proximal place away from the imaging site, a post-labeling delay time (PLD) is necessary to allow labeled spins reach the imaging site. If PLD is shorter than the arterial transit time (ATT), CBF underestimation will occur due to the insufficient arrival of the labeled spins.

ATT is the time of the labelled blood traveling from the labeling plane to the tissue voxel [3], which differs by regions and individual brains. By repeating data acquisitions at different PLDs, ATT can be estimated by fitting the ASL MRI perfusion signals at different PLDs to the kinetic model [4–6]. A practical issue, however, is the exponentially increased total scan time due to the repetitions of the entire imaging process at multiple PLDs. The long scan time not only increases the cost of imaging but also increases the risk of head motions, making the multiple PLD ASL MRI difficult to be implemented in clinical research[7, 8]. Reducing the number of PLDs will partially address the challenge but will make the ATT estimation less stable since the curve fitting-based estimation is prone to noise. There is a need to develop a method to reliably estimate ATT but without significantly increasing the total scan time, which however is highly challenging using conventional imaging method.

Prior-data guided deep machine learning can play a role in solving the above analytically impossible problem. Recent years, deep learning especially through the convolutional neural network (CNN) has achieved astonishing success in both high-level computer vision tasks such as image classification [9, 10], object detection [11] and low-level computer vision tasks such as image denoising [12] and image super-resolution [13]. Encouraged by the outstanding performance, deep learning has been introduced into many medical imaging processing fields as well [14, 15]. Some recent studies demonstrate the capability of deep learning in diverse medical image modalities, including MRI images [16–20], PET images [21–23], CT/X-ray images [24–26], ultrasound images [27], and CEST images [28–30]. The success inspired researchers to apply deep learning onto ASL MRI images. Deep learning offers advantages for denoising and artifact removal in ASL processing [31–34]. In addition, it helps to develop the parameter estimation models in ASL MR fingerprinting applications [35, 36]. For ATT and CBF estimation, deep learning also starts to play an important role. In [37], a 3D convolutional neural network (CNN) was proposed to estimate ATT and CBF simultaneously using PWIs at multi-PLDs. The same group further reduced the number of PLD acquisition without accuracy loss. Both experiments were conducted on 12 subjects [38]. In [39], another CNN and U-net were implemented to verify the feasibility of CBF and ATT estimation using deep learning for 1.5T and 3T multi-PLD ASL MRI data on 50 subjects, respectively. In this study, we proposed a novel network to significantly reduce the number of PLDs with higher SNR due to the CBF and ATT signal manifold learning from brain structure over 600 subjects.

Our contributions in this paper include: 1) explored the capability to learn a nonlinear transform from a few PWI images to both the CBF and ATT maps on a large-scale dataset; 2) improved the efficacy of CBF and ATT calculation by reducing the acquisition time by >60% compared to the conventional method; 3) increased the prediction image quality in terms of all the performance indices compared with the conventional method.

## 2 Method

### 2.1 Dataset Scheme

All the data was collected from Human Connectome Project aging dataset (HCP-A) [40]. The HCP-A protocol includes structural scans, task fMRI, resting state fMRI, diffusion, and cerebral blood flow (ASL), collected over two imaging sessions. Each session entails approximately 45 min of scanning, performed in a single day or across two days depending on site-specific procedures and constraints.

As detailed in the paper [41], the MRI protocol consisted of ASL was followed by Pseudo-continuous arterial spin labeling (PCASL) and 2D multiband (MB)-echo-planar imaging (EPI). ASL MRI was nearly included a 2D echo planar imaging acquisition, $86 \times 86 \times 65$ matrix, $3.5 \times 3.5 \times 3.5$ mm$^3$ voxel resolution, TR/TE = 8000/18.7 ms, pseudo-continuous labeling (1500 ms label duration, 5 PLDs of 200, 700, 1200, 1700, and 2200 ms collected in a sequential, non-interleaved fashion), and 6, 6, 6, 10, 15 control-label image pairs for each PLD, respectively. Imaging slice acquisition time was 0.054 ms. Two M0 images for CBF quantification were also acquired at the end of all the PLD acquisitions. To create the reference of CBF and ATT images, a voxel-wise ASL kinetic model fitting was applied using the PWIs from all the PLDs via FSL toolbox[43].

### 2.2 Network structure used in PLD2ATT

**Network Architecture** In order to validate the robustness of our model, one PLD and two PLDs were used as input for separate models. As shown in Figure 1, the one-PLD and two-PLD models have the same architecture inside. The DL based architecture is an improved deep residual network. The backbone of the network is the vanilla residual network while all the residual blocks are replaced by Wide-activation Deep Super-Resolution network (WDSR) blocks. The optimal number of WDSR blocks (8), the number of filters (32) and expansion ratio (4×) for WDSR filters are determined from the original WDSR paper [13]. The input for one-PLD model used the mean of the perfusion weighted image (PWI) normalized by M0 and the standard deviation of the corresponding PWIs for PLD=1.7s. The two-PLD model used the mean of the perfu-sion weighted image (PWI) normalized by M0 at PLD=0.7s and 1.7s and signal weighted delay [42] of these two PLDs as the input. To predict the CBF and ATT maps, two separate models shared the same input. Each input channel was associated with a separate convolutional layer. The output of all input layers was concatenated into one channel as the input to the successive layer. Following each convolutional layer were eight consecutive WDSR blocks. In each block, wide activation was used to retain the high-frequency tissue boundary information. Following the eight consecutive blocks, another convolutional layer without any activation function was attached to the end to get the output with additional input from the second layer. The output would generate the CBF and ATT maps, respectively. The reference of CBF and ATT were generated by FSL [43] using the PWIs from all the PLDs.

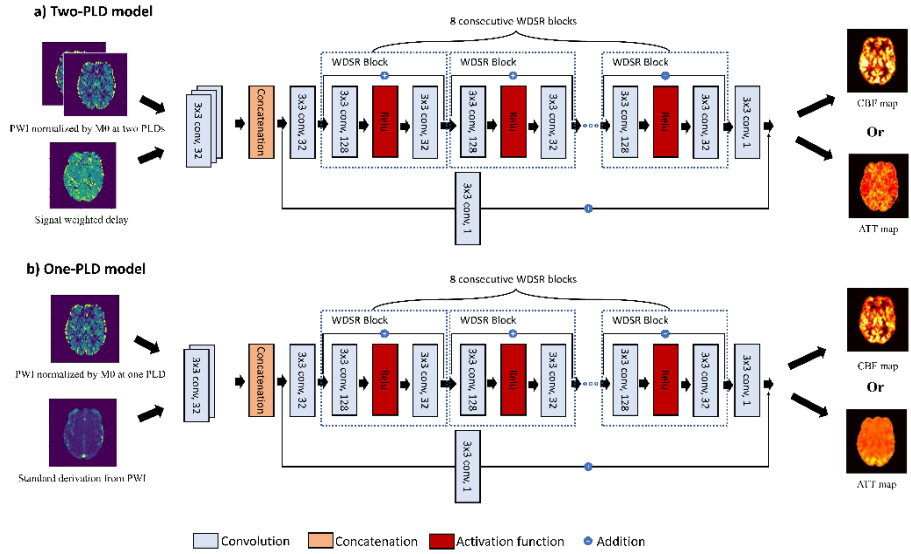

Figure 1 The architecture of the proposed method: a) two-PLD model and b) one-PLD model

**Training Details**. There are 500 subjects in the dataset. 300 of them are randomly selected for training, and another 100 subjects for validation and 100 subjects for testing. The image resolution is 86*86*60. For each subject, we extracted axial slices from slice 26 to slice 45 as the high-quality training data. With image size of 86*86 for each slice, totally 6000 image slices were used as the training set. All the training processes were implemented by using the entire image. Each experiment was run by 300 epochs and a batch size of 32. The mean absolute error (MAE) loss function was used in training. The ADAM optimizer with an initial learning rate of 0.001 was used to train the network. After 60 and 120 epochs, the learning rate was dropped by 0.1 and 0.05, respectively. All DL experiments were performed using framework of Keras and Tensorflow running on a Ubuntu18.04 system with NVIDIA GTX 2080ti.

## 3    Results

Visually, a couple of experiments were conducted to predict the CBF and ATT maps of all 100 subjects from testing set, respectively. Figure 2 showed an example of CBF estimation from one representative subject at different slices. As compared to the reference calculated by the conventional method in the first column, the deep learning methods showed CBF map quality improvement for two-PLD (the second column) and one-PLD (the third column) in fine details. The two models showed minor effects to the naked eye. The prediction was very consistent at different slices shown in each row.

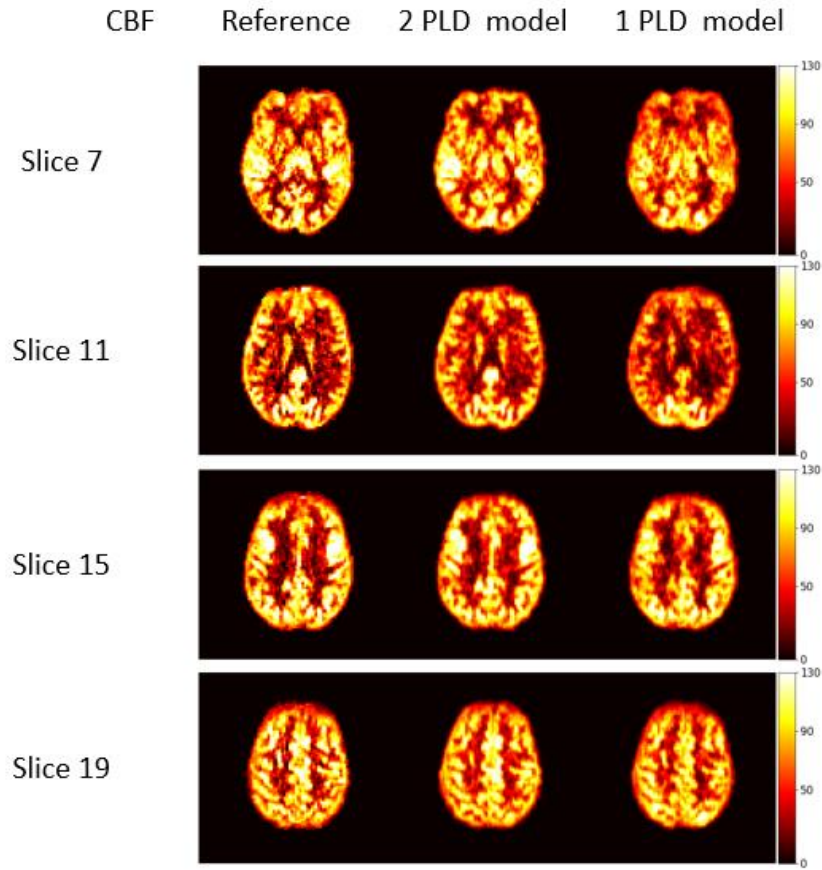

Figure 2 The visual results of the CBF maps of one representative subject at different slices by one-PLD and two-PLD models

Figure 3 demonstrated an example of ATT maps from one representative subject at different slices. The reference and the corresponding prediction from the two-PLD and one-PLD models were illustrated in the first, second, and third column, respectively. Unlike the CBF map, the difference between two-PLD and one-PLD models were nonnegligible on ATT estimation. Even both models showed higher SNR to the reference, the results from two-PLD model preserved structure much better than the ones from one-PLD model. The potential reason is that the reference of CBF has more stable structure compared to the reference of ATT, which makes the model learn the nonlinear transformation more accurately during the training process. More details come from the quantitative metrics.

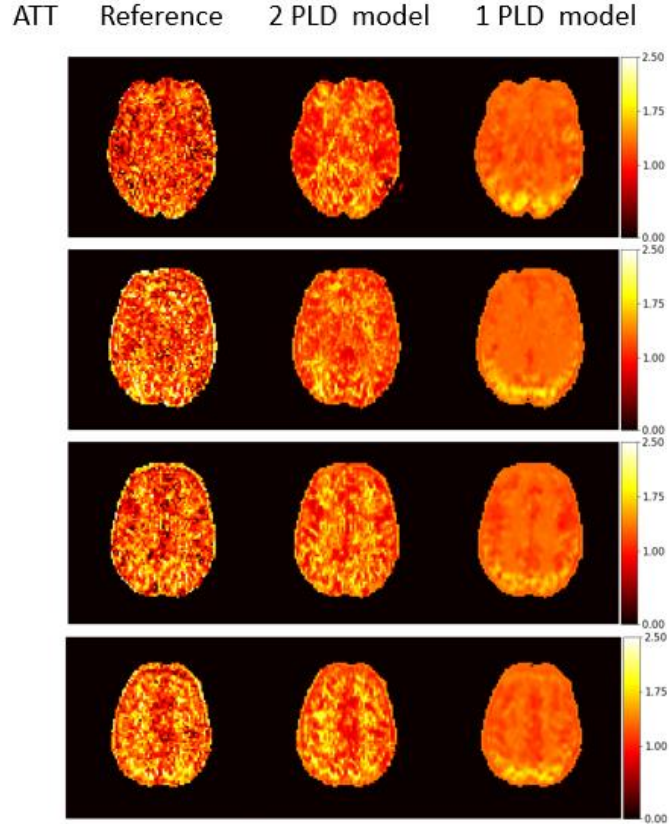

Figure 3 The visual results of the ATT maps of one representative subject at different slices by one-PLD and two-PLD models

Quantitatively, we calculated SSIM and PSNR to evaluate the capabilities of the proposed methods. Each slice of testing subject was evaluated by SSIM and PSNR separately. Table 1 shows the results of CBF maps from the two models. The two-PLD model is slightly better than one-PLD model in terms of SSIM and PSNR, but the difference is trivial. However, Table 2 shows similar results to our visual observations. The two-PLD model outperformed the one-PLD model significantly in terms of SSIM, which indicates the structure similarity. The PSNR values can barely see the difference as well as the two-PLD model.

Table 1 The SSIM and PSNR of CBF maps

| Method | SSIM | PSNR |
|---|---|---|
| one-PLD model | 0.83±0.062 | 26.56±2.19 |
| two-PLD model | 0.86±0.047 | 27.85±2.68 |

Table 2 The SSIM and PSNR of ATT maps

| Method | SSIM | PSNR |
|---|---|---|
| one-PLD model | 0.69±0.042 | 60.31±1.12 |
| two-PLD model | 0.80±0.039 | 61.48±2.10 |

## 4      Conclusion

In conclusion, we proposed a novel network to estimate the CBF and ATT maps using only one or two-PLD ASL MRI, which significantly accelerated the acquisition time and yield higher SNR than the conventional method.

**Acknowledgements.**    This   project   was   supported   by   some   NIH   grants: R01AG060054, R01 AG070227, R01EB031080-01A1, P41EB029460-01A1.